# 3rd International Symposium of the RGCS Network

**Arthur Sarazin, PhD Student**

University of Grenoble Alpes, CERAG Domaine universitaire - B.P. 47

38040 Grenoble Cedex 9 – France

arthur.sarazin@etu-iepg.fr

**Carine Dominguez-Péry, Professor**

University of Grenoble Alpes, CERAG Domaine universitaire - B.P. 47

38040 Grenoble Cedex 9 – France

carine.dominguez-pery@univ-grenoble-alpes.fr

**Khaled Bouabdallah, Professor**

University of Lyon, GATE

khaled.bouabdallah@universite-lyon.fr

# Open data ecosystems: what models to co-create service innovations in smart cities?

While smart cities are recently providing open data, how to organise the collective creation of data, knowledge and related products and services produced from this collective resource, still remains to be thought. This paper aims at gathering the literature review on open data ecosystems to tackle the following research question : what models can be imagined to stimulate the collective co-creation of services between smart cities' stakeholders acting as providers and users of open data ?

Such issue is currently at stake in many municipalities such as Lisbon which decided to position itself as a platform (O'Reilly, 2010) in the local digital ecosystem. With the implementation of its City Operation Center (COI), Lisbon's municipality provides an Information Infrastructure (Bowker et al., 2009) to many different types of actors such as telecom companies, municipalities, energy utilities or transport companies. Through this infrastructure, Lisbon encourages such actors to gather, integrate and release heterogeneous datasets and tries to orchestrate synergies among them so data-driven solution to urban problems can emerge (Carvalho and Vale, 2018). The remaining question being : what models for the municipalities such as Lisbon to lean on so as to drive this cutting-edge type of service innovation ?

## Open data ecosystems within smart cities

*From open data to open data ecosystems and platforms*

The idea of applying the metaphor of ecosystems to open data processing was first proposed by Pollock (2011), president of the Open Knowledge Foundation, a non-profit organisation promoting free access to all types of information, data included. He defined it by building on the "one way street" model where data providers publish it out into the world, then intermediaries such as software designers process it and eventually end users consume it. Data goes one way and one way only : from providers to users. Open data ecosystems add feedback loops to this model : intermediaires publish back the cleaned datasets they designed ; users report flagging errors back to providers.

Goëta (2016) came upon the same distinction by defining two types of systems related to open data. On the one hand, along with Pollock's one way street schema, open data is considered as a comprehensive initiative restricted to public actors aiming at releasing all the sets of *raw data* they produced (Ruppert, 2015 ; Zuiderwijk, 2015). On the other hand, open data is about selecting datasets based on their re-usability (defined as the potential to be used, at least once, after the free access provided by a smart city). From this approach, open data can be at the center of a dialogue between public entities and other actors playing the role of users (Denis and Goëta, 2013). Such dialogue implies interactions, feedback mechanisms and at a deeper level interdependencies between public entities and users, which reinforces the metaphor of open data cycle being an ecosystem. In parallel, this particular combination of collaboration among actors who can re-use data and technological elements enabling the transformation of open data gave birth to the concept of Open Government Data (OGD) Platform (Danneels et al., 2017)

*The distinct nature and properties of open data ecosystems*

Open data ecosystems actually cover distinct features. Open data ecosystems may be seen as "intentionally cultivated for the purpose of achieving (the) managerial and policy vision (of public entities)" (Harrison et al., 2012).

Yet, this public-centric approach fails to grasp the complexity of the entanglement between actors involved in the production and use of open data and calls for new ways of organising between the three main actors and the open data platform to co-create services : public entities, technological innovators and users. (Figure 1.).

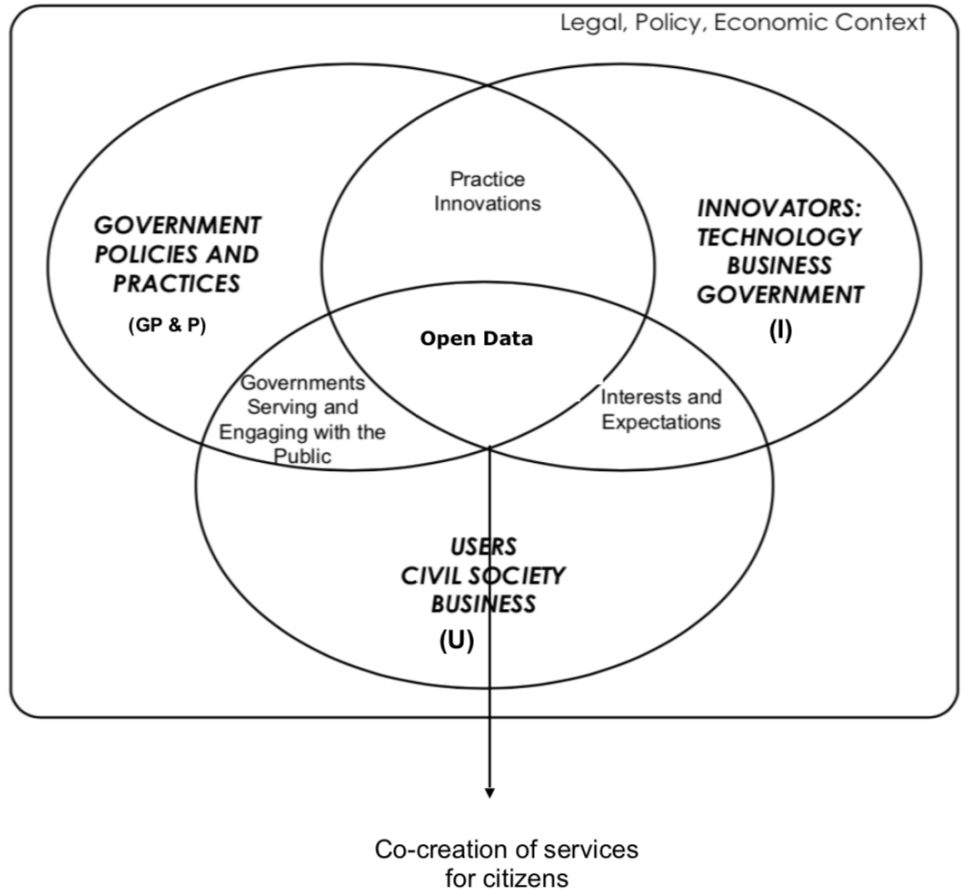

*Figure 1 : Categories of actors involved in open data ecosystems (adapted from Harrison et al., 2012 and Ruijer et al., 2017)*

This organisational issue is on the three actors' agenda since the European Commission (2006) forecasted a 40€ billion market related to open data re-use. It is notably raising awareness among smart cities development projects. Indeed, following Boullier (2016), the shape and management of smart cities depends heavily on the way urban data production, storage and use is distributed among smart cities' stakeholders (Table 1.)

| The "good old city" | Data remains in the traditional silos, each producer of data remaining the exclusive owner of it |
|---|---|
| The "IBM city" | A city driven integrally from a centralized station capturing all data flows and establishing dashboards that will make the most rational decisions for the operation of the city. |
| The "Google city" | All the data produced in the urban space are connected without presuming any use beforehand. This form refers to an increased use of big data in the management of cities. |
| The "Wiki city" | The city is based this time on collecting contributions from citizens. A form therefore where the place of *crowdsourcing* is preponderant to make decisions about the functioning of the city. |

Table 1. Typology of smart cities based on alternative ways of managing urban data. From Boullier (2016)

# Service innovation in open data ecosystems: from data providers to data users

Service innovation in open data ecosystems describe the fact that data providers give data to users, so they can co-create services toward collective interests (Attour and Rallet 2014). More precisely, it breaks down into 3 elements (Lusch and Nambisan, 2015) : the open data ecosystem, the open data platform and value co-creation which refers to "the processes and activities that underlie resource integration and incorporate different actor roles in the ecosystem" (Lusch and Nambisan, 2015, p. 162). In this paper and the related PhD research, we decided to focus on value co-creation mechanisms incorporating two actor roles : open data providers and open data users.

As an example, open data service innovation processes can be observed in data mash-up services such as the site WhereDoesMyMoneyGo.com that permit citizens to understand how public money is distributed by mixing datasets coming from different providers.

*Perspective on open data provider's side* : *promises and challenges*

Open data providers are actors responsible of "the nature and quality of open data sets: their legal, technical and social openness, relevance and quality".
Service innovation holds many promises for them. Public actors might observe an improvement in democratic governance, political participation and public service efficiency. Technological innovators might benefit from users' feedback and improve their data-driven processes, products and services resulting in a surplus of growth and competitiveness (Dawes et al., 2016)
However, open data providers face high barriers when engaging in service innovation processes. Firstly, they may be asked by users to release datasets that are politically sensitive or that might require some technical investment they are not able to take (Davies, 2014). Secondly, they may not be able to guarantee the sustainability of the datasets to data users. Thirdly, they might be overwhelmed with the increasing amount of data being available in several locations (Eckelberg, 2018) and struggle with complex and multi-actor governance mechanisms (Bauer & Kaltenbock, 2012)

*Perspective on open data user's side: promises and challenges.*

By users, we mean the technology-based and virtual communities (Rheingold, 1985) that are in capacity to make direct use of open data. When engaged in services innovation, they benefit from resources (data, funding, knowledge) they can leverage to cater their specific needs (Attour and Rallet, 2014). Users involved regularly might even structure themselves as a consequence of the repeated exchanges with institutionalized actor and of the underlying resource integration service innovation implies (Lusch and Nambisan, 2015). From this idea Bourcier (2013) lays down the hypothesis that open data users, when adopting an organization similar to virtual communities (Rheingold, 1985), can become a viable alternative to the private sector in smart cities and to a greater extent, co-regulators of public action (Bourcier, 2013). Such communities could certainly produce products and services by sourcing resources, knowledge and skills from their own members, a production model similar to the commons-based peer production model invented by Benkler (2009). As to co-regulating public action, Bourcier (2013) emphasizes that, when involved in service innovation, the governance rules virtual communities created for themselves complement public sector rules.

Still, co-innovation biggest challenge for users is to trigger an incremental dynamic where already existing services are constantly upgraded through the active participation of several types of actors with no clear business models. (Yu, 2016)

**Models of open data interactions within open data ecosystems**

Interactions between the different actors have yet only been studied at a generic level. For example, McLeod and Naughton (2016) approached Open Data Ecosystem with Actor-Network Theory (ANT) and generalised interactions between ecosystem actors as "arrangements". Zuiderwijk (2015) focused on user's interactions, putting aside the issue of their interaction with data providers. Kuk and Davies (2011) proposed a scheme embracing both open data providers and open data users, but conceived suppliers as an input in a waterfall process and users as active forces contributing to the unfolding and achievement of the process, without describing their relationships. Finally, Pollock (2011) compared the characteristics of such dialogue to an ecosystem in a socio-technical domain, where both technological elements and actors interact in a tightly manner with substantial interdependencies (Harrison et al., 2012).

To complete these theoretical backgrounds, we can draw parallels with open source's activists who borrowed some of its principles (Heimstädt et al., 2018), such as the emphasis on re-using existing artefacts (Haefliger et al., 2008), the freedom to contribute or to quit a project, and collaborative practices where users take an active part in the design of artefacts (Raymond, 1999). In addition, the framework of knowledge commons (Hess and Oström, 2007) was used to understand open data ecosystems (Dawes et al., 2016). Indeed, open datasets can be considered as knowledge commons due to the license terms under which they are released and because any actor is free to participate in the management of such resource. However, contrary to pure knowledge commons, the governance rules around such resource are not self-managed by a community of users.

Eventually, we argue that these ecosystems are a one-of-a-kind, laying at the intersection between government platforms ecosystem (0'Reilly, 2011), and a bazaar model that is " a collection of selfish agents attempting to maximize utility which in the process produces a self-correcting spontaneous order" (Raymond, 1999). A theoretical perspective needs to be developed to apprehend the peculiarity of such ecosystems.

# The models to stimulate service innovation in smart cities : a wicked problem to tackle through a Design Science Research Methodology (Hevner et Chatterjee, 2010)

Among the factors explaining the peculiarity of such ecosystems stands the context in which we decided to study them : smart cities. As Dameri (2017) puts it, smart city is a "x-city" notion where the "x takes a different meaning based on the dimensions considered of high priority by a territory" (Silva-Morales, 2017). Such conceptualization has 2 main implications that will guide the methodology we will unfold in the PhD research related to this paper.

First, smart city depends on what a territory has considered as high priority, which means that smart city *is not* a tangible and objective phenomenon but rather the embodiment of an idea of what a city *should be* or, as Picon (2013) puts it, a self-fulfilling ideal. As a consequence, our research problem is inextricably, inherently goal-oriented and calls for solutions that will assist cities in reaching their ideal.

Karl Popper argues in *The logic of scientific discovery* that it is a principle of science that solutions to problems are only hypotheses offered for refutation, each refutation being a step toward truth. We can clearly see that our research problem does not fit into this conception of science as its solution will not be hypotheses standing in the quest of truth but a way to improve some characteristics of the cities where people live (Rittel et Weber, 1973).

Second, this conceptualization implies that the research process and the solutions we might sketch will be highly dependent on the smart cities' stakeholders subjectivity. Depending on the dimensions they claim to be of prime interest, the answers will change. For example, it means that in case of a changing political landscape, the service innovation phenomenon observed might suddenly stop because the new political leaders redefined the smart city notion and re-allocated the funds for service innovation initiatives.

Taken together, those 2 points indicates that our research problem is a *wicked problem* (Rittel et Weber, 1973) which calls for a peculiar methodology : the Design Science Research Methodology (Hevner and Chatterjee, 2010). Its peculiarity mainly originates from the fact that the final results include, aside scientific publications, the design of an artifact that can be tested by the practitioners of a field.

**The Design Science Research Methodology (DSRM) to be developed**

Following Hevner and Chatterjee (2010), our methodology breaks down into five phases (Figure 2.)

| Number and name of the phase | Related research activities |
|---|---|
| 1-Awareness of problem | Locate a specific lack of knowledge (SLOK) in the area of interest that is shared between the research community and the practitioners |
| 2-Suggestion | Refine the SLOK to one or more research question by : <ul><li>elaborating the unknown factors</li><li>reviewing applicable research techniques</li><li>determining the interest of the question to the topic research community</li><li>determining publish-ability</li><li>Scoping to research community standards and resource limitations</li></ul> |
| 3-Development | Develop an artifact through prototyping and evaluate its impact on practitioners |
| 4-Evaluation | |
| 5-Publication | Write up the results and publish |

To conduct the first phase (e.g Awareness of the problem phase) we seized the opportunity offered by a well established open data intermediary in the french open data ecosystem to follow their activities with open data providers, open data users and other intermediaries. Acting as a mediator between providers and users to make open data sets useful, they have been offering us access to their entire network. So far, we have observed and participated in dozens of meetings and phone calls with 4 municipalities, 4 technological innovators in the private sector and 7 intermediaries including open data platform developers and data mediators developers. We also conducted 7 exploratory interviews with 1 municipality, 1 technological innovator and 4 intermediaries.

The substantial amount of data gathered will be analysed using grounded theory (Strauss and Corbin, 1990) that "requires not only that data and theory be constantly compared and contrasted during data collection and analysis but also that materializing theory drives ongoing data collection" (Locke, 1996). That is we will categorize data, compare them across observations in order to create theoretical statements on models. These statements will then drive deeper data collection through qualitative case studies of 2 to 3 smart cities (e.g Suggestion phase of the DSRM). Following the same data analysis pattern, these case studies will help us refine the theoretical statements on models and propose alternative theoretical perspectives to be tested through the Development and Evaluation of an artifact (e.g 4th and 5th phase of the DSRM). It will permit us to refine empirically the models so they can become the "solutions to real-world problems of interest to practice" that are the ultimate goals of our DSRM methodology.

All in all, open data ecosystems are raising fascinating questions related to (co-)creation mechanisms toward service innovation and call for innovative scientific methodologies. We wish to have the opportunity to share the state of the art on open data ecosystems at RGCS Barcelona, launch conversations about Design Science Research Methodologies and complete our understanding toward other smart cities collective co-creation contexts.